\documentclass[aps,twocolumn,showpacs,pra,twoside,amssymb,amsmath]{revtex4}
\usepackage{amssymb}
\usepackage{graphicx}
\usepackage{amsmath}
\usepackage{colordvi}
\usepackage{bbm}
\usepackage{amsmath}

\newcommand{\idol}{\ensuremath{\mathbbm 1}}
\newcommand{\tr}{{\rm Tr}}
\newcommand{\rank}{{\rm rank}}

\begin{document}

\title{Detecting the quantum discord of an unknown state by a single observable}

\author{Chengjie Zhang$^{1}$}
\email{cqtzcj@nus.edu.sg}
\author{Sixia Yu$^{1,2}$}
\author{Qing Chen$^{1,2}$}
\author{C.H. Oh$^{1}$}
\email{phyohch@nus.edu.sg}
\affiliation{$^1$Department of Physics and Centre for Quantum Technologies, National University of Singapore, 117542, Singapore \\
$^2$Hefei National Laboratory for Physical Sciences at Microscale and Department of Modern Physics, \\ University of Science and Technology of China, Hefei, Anhui 230026, China}

\begin{abstract}
We propose a single observable to witness the nonzero quantum discord of an unknown quantum state provided that we have four copies of the state. The expectation value of this observable provides a necessary and sufficient condition for nonzero quantum discord in $2\times N$ systems and a necessary condition in higher finite-dimensional bipartite systems. Furthermore, a nontrivial lower bound of quantum discord can be obtained from this expectation value. The proposed observable can be experimentally measured in exactly the same easy manner as that of the entanglement witness, moreover a quantum circuit is designed to determine the expectation value of our observable with four simultaneous local qubit-measurements.
\end{abstract}

\pacs{03.67.-a, 03.65.Ta, 03.67.Lx}

\maketitle
\section{Introduction}
Entanglement is an essential resource in almost all quantum computing and informational processing tasks. In certain quantum computing tasks, however, there are still quantum advantages with the absence of entanglement. One typical example is the deterministic quantum computation with one qubit (DQC1) \cite{DQC1} in which the {\it quantum discord}, introduced by Ollivier and Zurek \cite{discord1} and independently by Henderson and Vedral \cite{discord2}, is proposed to be responsible for the quantum speedup \cite{Datta}. Ever since, the quantum discord, a nonnegative number quantifying all the quantum correlations in a state beyond the entanglement, has attracted much interest in quantum information theory \cite{Modi,Lang,phase,gaussian,Markovian1,Markovian2,Markovian3,Markovian4,Markovian5}. Subsequently the operational meaning of the quantum discord has established firmly its status as an essential resource \cite{merge}.

Quantum discord provides a measure for the quantum correlation beyond entanglement, which is defined as
\begin{equation}\label{DA}
    D_A(\varrho)=\min_{\{E_k\}}\sum_k p_k S(\varrho_{B|k})+S(\varrho_A)-S(\varrho),
\end{equation}
where $S(\varrho)=-\mathrm{Tr}(\varrho\log_2\varrho)$ is the von Neumann entropy, $\varrho_A$ ($\varrho_B$) is the reduced density matrix of subsystem $A$ ($B$), and the minimum is taken over all possible positive operator-valued measures $\{E_k\}$ on subsystem $A$ \cite{discord2} with $p_k=\mathrm{Tr}(E_k\otimes\idol\varrho)$ and $\varrho_{B|k}=\tr_A(E_k\otimes\idol\varrho)/p_k$. The minimum can also be taken over all the von Neumann measurements \cite{discord1}, and these two definitions coincide in the case of zero quantum  discord. In general, the quantum discord  is asymmetric, i.e., $D_A(\varrho)$ is not equal to $D_B(\varrho)$ which is obtained by measuring subsystem $B$. For simplicity we only consider $D_A(\varrho)$ in the following.

Though almost all quantum states have nonzero quantum discord \cite{almost}, states with vanishing quantum discord find numerous applications in fundamental concepts in addition to its initial motivation in pointer states \cite{discord1}. For examples, the vanishing quantum discord is related to the complete positivity of a map \cite{CP} and the local broadcasting of the quantum correlations \cite{broadcast,luo}. For a given bipartite state in a finite-dimensional system, sufficient conditions as well as necessary and sufficient conditions for nonzero quantum discord have been proposed \cite{2N,condition1,condition2,condition3} in the form of local commutativity, strong positive partial transpose, etc. Specially Rahimi and SaiToh  \cite{Rahimi} have introduced an example of nonlinear witness for the nonzero quantum discord. All these criteria, however, require some knowledge of the quantum states and have nothing to say about the nonzero values of the quantum discord. The problem of witnessing the nonzero quantum discord of an unknown state remains open.


In this paper we shall propose a single observable to detect the nonzero quantum discord of a completely unknown state provided that we have four copies of the state. The expectation value of this observable provides not only a necessary and sufficient condition for nonzero quantum discord in $2\times N$ systems and a necessary condition in higher finite-dimensional bipartite systems, but also quantitative lower bounds for the quantum discord and the geometric measure of quantum discord. Our observable can be experimentally measured in exactly the same easy manner as that of the entanglement witness, and moreover a quantum circuit is designed so that its expectation value can be easily determined via four simultaneous local qubit-measurements.

\section{Single-observable detection of quantum discord}
It turns out that, similarly to entanglement witnesses, there is a Hermitian observable $W$ that detects perfectly the nonzero quantum discord without state tomography and provides a lower bound for the quantum discord. To construct this observable we note firstly an important lesson learned from \cite{Rahimi} that a linear witness can never detect nonzero discord of a separable state. Therefore nonlinear witnesses should be used to detect quantum discord and the expectation values of an observable on multi copies of a state is naturally nonlinear. Secondly, since the quantum discord is invariant under local unitary (LU) transformations, the observable witnessing the quantum discord should also be LU-invariant.

In general, the polynomial LU invariants of degree $k$ for a bipartite given quantum state $\varrho_{AB}$ is given by $\mathrm{Tr}(U\varrho^{\otimes k})$, where $U$ is a tensor product of some permutation operator acting on copies of subsystem $A$ and some permutation operator of subsystem $B$ \cite{invariants}. For later use we introduce the following four LU invariants $\mathrm{Tr}(U_i\varrho^{\otimes 4})$ $(i=1,2,3,4)$ of order 4 where
\begin{eqnarray}\label{U}
\begin{aligned}
&&U_1=V_{14}^{A}V_{23}^{A}V_{12}^{B}V_{34}^{B},\quad
U_2=V_{14}^{A}V_{12}^{B}V_{34}^{B},\\
&&U_3=V_{12}^{A}V_{34}^{A}V_{12}^{B}V_{34}^{B},\quad
U_4=V_{12}^{A}V_{12}^{B}V_{34}^{B},
\end{aligned}
\end{eqnarray}
in which we have denoted by $V_{ij}^{A,B}=\sum_{kl}|kl\rangle\langle lk|_{ij}^{A,B}$ the swapping operator acting on the $i$-th and the $j$-th copies of subsystem $A$ ($B$).  Our discord witness reads
\begin{equation}\label{W}
 W=U_1-U_3-\frac 2{d_A}(U_2-U_4),
\end{equation}
where $d_A$ is the dimension of $\mathcal{H}_A$ and obviously $\mathrm{Tr}(W\varrho^{\otimes4})$ is also an LU-invariant. We are now ready to present our main result with the proof shown in the appendix:

\textit{Theorem.} For an unknown bipartite state $\varrho$  of a $d_A\times d_B$ system with 4 copies, if $\mathrm{Tr}(W\varrho^{\otimes4})=0$ then $D_A(\varrho)=0$, conversely if $D_A(\varrho)=0$ then $\mathrm{Tr}(W\varrho^{\otimes4})=0$ when $d_A=2$. Moreover, if one restricts to von Neumann measurements in Eq. (\ref{DA}),
\begin{eqnarray}\label{lower}
D_A(\varrho)\geq\max\{0, S(\varrho_A)-S(\varrho)-\log_2[d_AQ(\varrho)]\}
\end{eqnarray}
can be obtained, where
\begin{eqnarray}
Q(\varrho):=\frac{\tr\varrho_B^2}{d_A}+(d_A-1)\sqrt{\tr( W\varrho^{\otimes4})+(\tr\tilde\varrho^2)^2}
\end{eqnarray}
with $\tilde\varrho:=\varrho-\idol_A/d_A\otimes\varrho_B$.

\textit{Remark 1.} The expectation value of $W$ provides not only a necessary and sufficient condition for nonzero quantum discord in $2\times N$ systems, but also a necessary condition in $M\times N$ systems with $M>2$.

\textit{Remark 2.} Similarly to Eq. (\ref{lower}), one can also present a nontrivial lower bound for the geometric measure of quantum discord using the expectation value of $W$, i.e.,
\begin{eqnarray}
D_A^{(2)}(\varrho)\geq\max\{0,\tr\varrho^2-Q(\varrho)\}.\label{lower2}
\end{eqnarray}
$D_A^{(2)}(\varrho)$ is the geometric measure of quantum discord defined as \cite{condition1}
\begin{equation}\label{}
    D_A^{(2)}(\varrho)=\min_{\chi\in\Omega_0}\|\varrho-\chi\|^2,
\end{equation}
where $\Omega_0$ denotes the set of zero-discord states and $\|X-Y\|^2=\tr[(X-Y)^2]$ is the square norm in the Hilbert-Schmidt space.

\textit{Remark 3.} Suppose that $\{A_i\}_{i=0}^{d^2_{A}-1}$ and $\{B_j\}_{j=0}^{d^2_{B}-1}$ are complete sets of local orthogonal observables (LOOs) for subsystem $A$ and $B$ \cite{LOO}, respectively, which satisfy $\mathrm{Tr}(A_i A_j)=\mathrm{Tr}(B_i B_j)=\delta_{ij}$ with $A_0=\idol^A/\sqrt{d_A}$ and $B_0=\idol^B/\sqrt{d_B}$. Denote that $r_{ij}=\langle A_i\otimes B_j\rangle_{\varrho}$, and $R:=(r_{ij})$ for $i=1,\cdots,d_A^2-1$ and $j=0,\cdots,d_B^2-1$ is a $(d_A^2-1)\times(d_B^2)$ dimensional matrix. Thus,
\begin{equation}\label{rho}
    \varrho=\sum_{i=1}^{d_A^2-1}\sum_{j=0}^{d_B^2-1}r_{ij}A_i\otimes B_j+\frac{\idol^A}{d_A}\otimes\varrho_B.
\end{equation}
It can be obtained that $\mathrm{Tr}(RR^T)=\sum_{i=1}^{d_A^2-1}\sum_{j=0}^{d_B^2-1}\langle A_i\otimes B_j\rangle_{\varrho}\langle A_i\otimes B_j\rangle_{\varrho}=\mathrm{Tr}(\bar{V}_{12}^A V_{12}^B \varrho^{\otimes 2})$, where we have used $V^B=\sum_{j=0}^{d_B^2-1}B_j\otimes B_j$ and defined $\bar{V}^{A}:=\sum_{i=1}^{d_A^2-1}A_i\otimes A_i=V^{A}-\idol^{A}\otimes\idol^{A}/d_A$. Similarly, we have $\mathrm{Tr}[(RR^T)^2]=\sum_{i,k=1}^{d_A^2-1}\sum_{j,l=0}^{d_B^2-1}\langle A_i\otimes B_j\rangle_{\varrho}\langle A_k\otimes B_j\rangle_{\varrho}\langle A_k\otimes B_l\rangle_{\varrho}\langle A_i\otimes B_l\rangle_{\varrho}=\mathrm{Tr}(\bar{V}_{14}^{A}\bar{V}_{23}^{A}V_{12}^{B}V_{34}^{B}\varrho^{\otimes 4})$. Therefore,
\begin{eqnarray}\label{R}
\mathrm{Tr}[(RR^T)^2]-[\mathrm{Tr}(RR^T)]^2=\mathrm{Tr}(W'\varrho^{\otimes4})=\mathrm{Tr}(W\varrho^{\otimes4}),
\end{eqnarray}
where $W'=\bar{V}_{14}^{A}\bar{V}_{23}^{A}V_{12}^{B}V_{34}^{B}-\bar{V}_{12}^{A}\bar{V}_{34}^{A}V_{12}^{B}V_{34}^{B}$ and the second equality holds since the invariance of $\varrho^{\otimes4}$ under arbitrary permutation of different copies. From Eq. (\ref{R}), one can see that $W$ is related to the rank of $R$ and
the expectation value of $W$ is always non-positive for arbitrary states.

Let us take the output state of DQC1 model as an example.
The model of DQC1 introduced by Knill and Laflamme can evaluate
the normalized trace of a unitary matrix efficiently \cite{DQC1}.
Initially there is a single control qubit in the state $(\idol+\alpha\sigma_z)/2$ with $\alpha$
describing the purity of the control qubit and a collection of $n$ qubits in the completely mixed state $\idol_n/2^n$.
After the Hadamard gate applied to the control qubit and a control $n$-qubit unitary gate $U_n$, the output state becomes
\begin{eqnarray}
\varrho_{out}=\frac{1}{2^{n+1}}
\left(\begin{array}{cc}
 \idol_n & \alpha U_{n}^{\dag} \\
 \alpha U_{n} &  \idol_n
\end{array}\right).\label{final}
\end{eqnarray}
By making the measurements in the eigenbasis of $\sigma_x$ and $\sigma_y$ on the control qubit, the quantum circuit provides the normalized trace of $U_n$, $\tau=\mathrm{Tr}(U_n)/2^n$, since the expectation values of $\sigma_x$ and $\sigma_y$ give estimates for the real and imaginary parts of $\tau$, respectively.

There is strong evidence that DQC1 task cannot be simulated efficiently using classical computation \cite{evidence}, and the quantum discord has been proposed as a figure of merit for characterizing the resources present in this computational model \cite{Datta}. It is worth noticing that the output state (\ref{final}) is a natural $2\times N$ state under the bipartite split between the control qubit and the collection of $n$ qubits. Therefore, we can derive an explicit condition for characterizing the correlations in the output state using the theorem. After some algebra, we have
\begin{equation}\label{wdqc1}
    \mathrm{Tr}(W\varrho_{out}^{\otimes4})=\frac{\alpha^4}{2^{2n+3}}\bigg(\bigg|\frac{\mathrm{Tr}U_n^2}{2^n}\bigg|^2-1\bigg).
\end{equation}
According to our theorem, $D_A(\varrho_{out})=0$ if and only if $\mathrm{Tr}(W\varrho_{out}^{\otimes4})=0$, i.e., $\alpha=0$ or $U_n^2=e^{i\phi}\idol_n$ for some real $\phi$, which is equivalent to the result shown in Ref. \cite{condition1}.

On the other hand, if $\mathrm{Tr}U_n=0$ and $\alpha=1$, using Eq. (\ref{lower}) we have a lower bound
\begin{equation}\label{dqc1lower}
    D_A(\varrho_{out})\geq 1- \log_2\bigg(1+\sqrt{2^{2n+2}\mathrm{Tr}(W\varrho_{out}^{\otimes4})+1}\bigg),
\end{equation}
for the quantum discord in DQC1. It is obvious that the negative expectation value $\mathrm{Tr}(W\varrho_{out}^{\otimes4})$ provides a nontrivial lower bound for $D_A(\varrho_{out})$. It has been shown that for almost every unitary matrix $U_n$ (random unitary) the discord in the output state (\ref{final}) is nonvanishing \cite{Datta}. However, in the nontrivial case $U_n^2=e^{i\phi}\idol_n$ the quantum discord $D_A(\varrho_{out})=0$, which provides a counter example for the nonvanishing quantum discord of every nontrivial unitary matrix $U_n$. Thus, the quantum discord might not be the reason for the quantum speedup in DQC1.

\begin{figure}
\begin{center}
\includegraphics[scale=0.65]{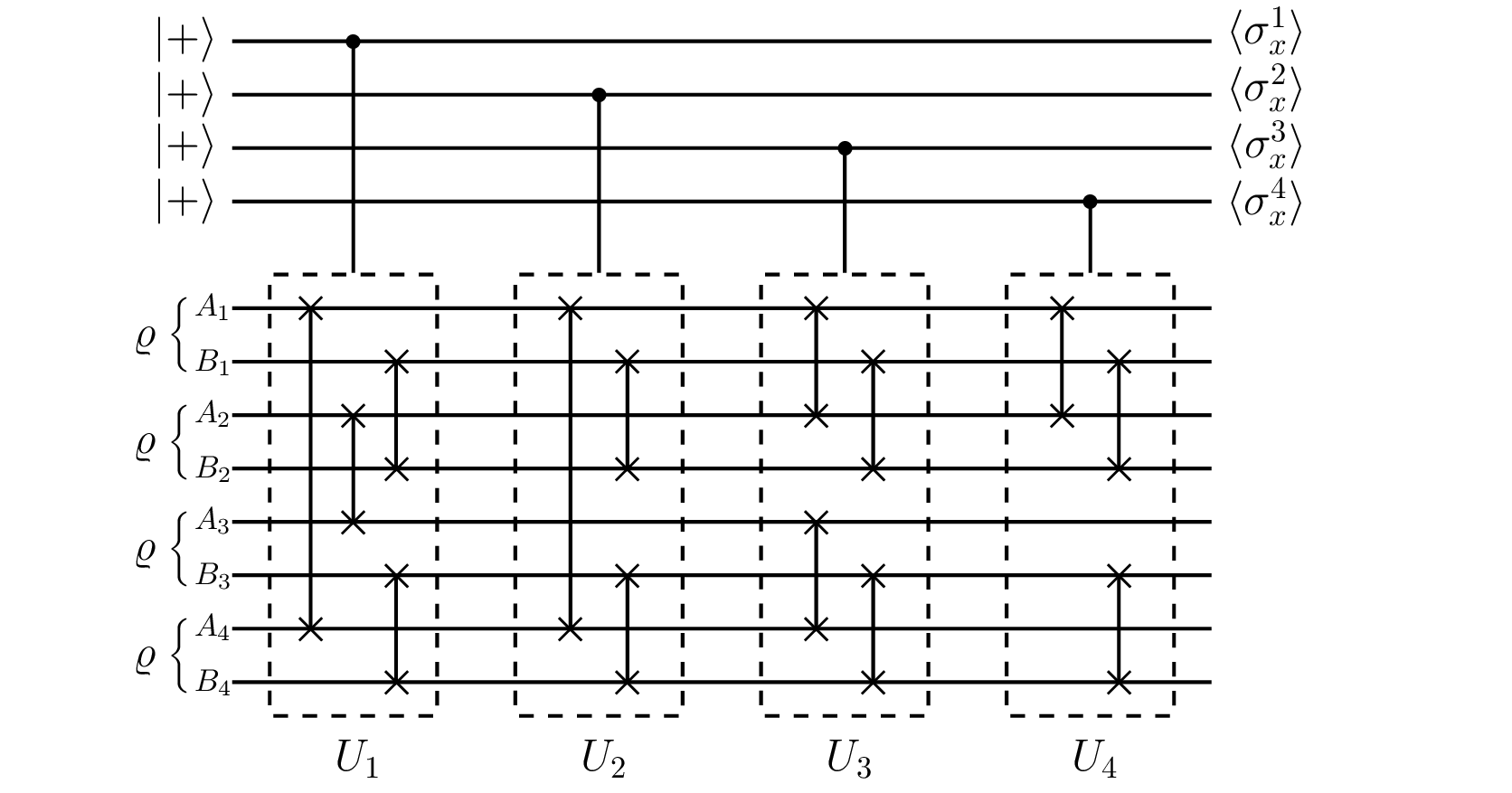}
\caption{Quantum circuit to measure the quantum discord witness $W$.
Four auxiliary qubits, initially prepared in state $|+\rangle$, are control qubits for controlled-$U_i$ ($i=1,2,3,4$) operations. Each $U_i$ is composed of several controlled-swapping gates which are represented by connected crosses. After all the controlled-$U_i$ operations, we perform $\sigma_x$-measurements on the auxiliary qubits and via Eq. (\ref{sigma}) in the text value of $\mathrm{Tr}(W\varrho^{\otimes 4})$ can be obtained.}\label{1}
\end{center}
\end{figure}

\section{Measuring the discord witness $W$}
In order to measure the expectation value of the observable $W$ we have designed a quantum circuit as shown in Fig. \ref{1}. We have introduced four auxiliary qubits, which are all initially prepared in the state $|+\rangle=(|0\rangle+|1\rangle)/\sqrt{2}$. After the controlled-$U_i$ ($i=1,\cdots,4$) operations with each qubit as source, we preform $\sigma_x$-measurements on all four auxiliary qubits so that four expectation values are obtained simultaneously. As it turns out
$\langle \sigma_x^1\rangle=\mathrm{Tr}(U_1\varrho^{\otimes4})$,
$\langle \sigma_x^2\rangle=\mathrm{Tr}(U_2\varrho^{\otimes4})$,
$\langle \sigma_x^3\rangle=\frac12(\mathrm{Tr}(U_3\varrho^{\otimes4})+\langle \sigma_x^1\rangle)$, and
$\langle \sigma_x^4\rangle=\frac12(\mathrm{Tr}(U_4\varrho^{\otimes4})+\langle \sigma_x^2\rangle)$,
from which one can easily calculate the value of $\mathrm{Tr}(W\varrho^{\otimes 4})$ according to
\begin{equation}\label{sigma}
    \mathrm{Tr}(W\varrho^{\otimes 4})=2\Big(\langle \sigma_x^1\rangle-\langle \sigma_x^3\rangle\Big)-\frac{4}{d_A}\Big(\langle \sigma_x^2\rangle-\langle \sigma_x^4\rangle\Big).
\end{equation}

\begin{figure}
\begin{center}
\includegraphics[scale=0.47]{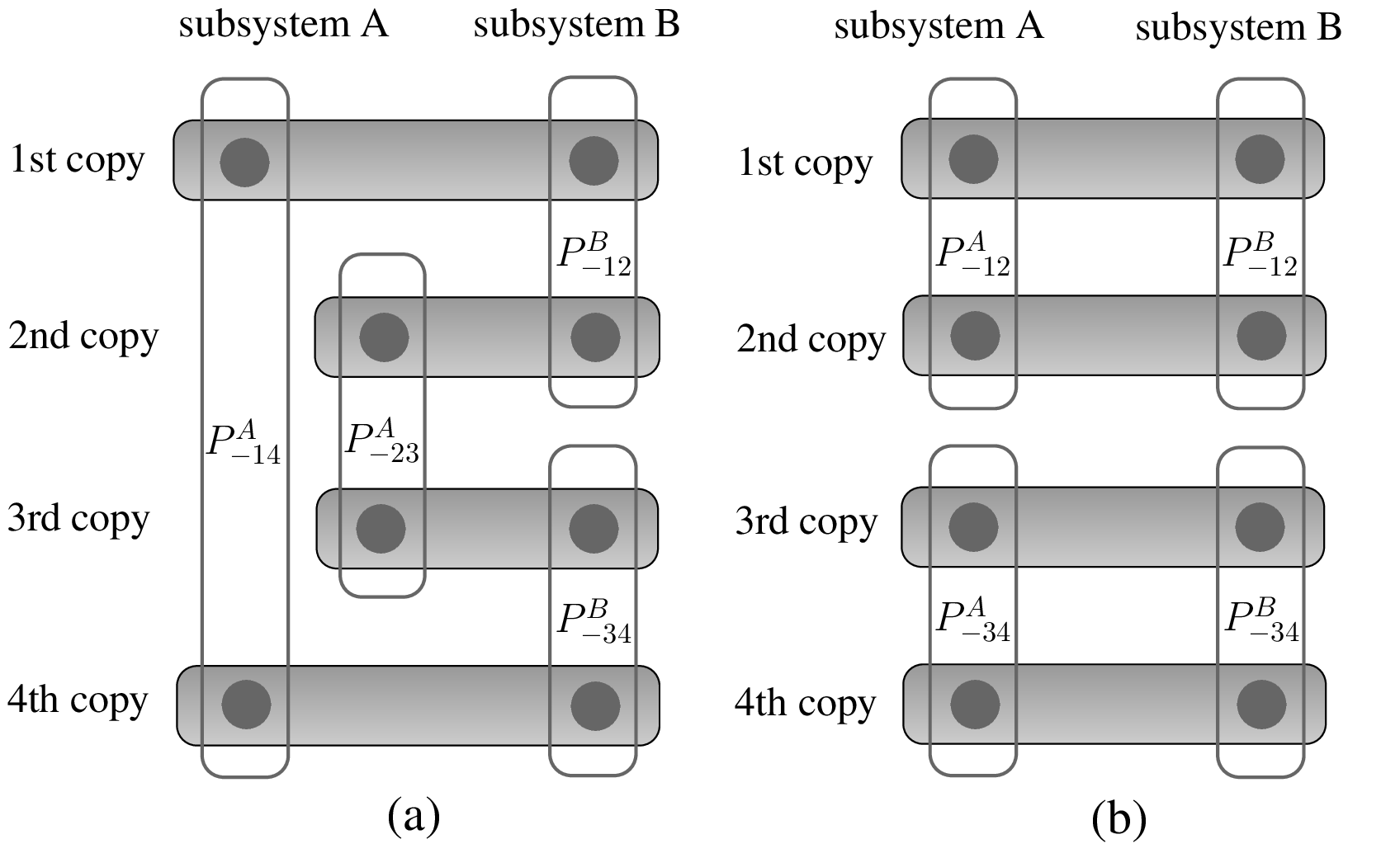}
\caption{Illustration of direct measurement of the observable $W$ using four copies of $\varrho$. There are two measurement settings $P_{-14}^{A}P_{-23}^{A}P_{-12}^{B}P_{-34}^{B}$ and $P_{-12}^{A}P_{-34}^{A}P_{-12}^{B}P_{-34}^{B}$ corresponding to (a) $U_1$ and $U_2$, and (b) $U_3$ and $U_4$.}\label{0}
\end{center}
\end{figure}

Besides the quantum circuit shown above, one can also directly measure the single observable without the quantum circuit. Let $P_-=(\idol-V)/2$ be the antisymmetric projector on some two copies and we have $V=\idol-2P_-$ so that
we can expand $W$ with $P_{-ij}$ as what follows
\begin{eqnarray}
W&=&(\idol_{14}^{A}-2P_{-14}^{A})\otimes\left(\frac{d_A-2}{d_A}\idol_{23}^{A}-2P_{-23}^{A}\right)\nonumber\\
&&\otimes(\idol_{12}^{B}-2P_{-12}^{B})\otimes(\idol_{34}^{B}-2P_{-34}^{B}) \nonumber\\
&&-(\idol_{12}^{A}-2P_{-12}^{A})\otimes\left(\frac{d_A-2}{d_A}\idol_{34}^{A}-2P_{-34}^{A}\right)\nonumber\\
&&\otimes(\idol_{12}^{B}-2P_{-12}^{B})\otimes(\idol_{34}^{B}-2P_{-34}^{B}).\label{P_W}
\end{eqnarray}
Since the measurement of the observable $P_{-14}^{A}P_{-23}^{A} P_{-12}^{B} P_{-34}^{B}$ gives also information about observables like $P_{-23}^{A} P_{-12}^{B}P_{-34}^{B}$, etc.,  we need  only two measurement settings $P_{-14}^{A} P_{-23}^{A} P_{-12}^{B} P_{-34}^{B}$ and $P_{-12}^{A} P_{-34}^{A} P_{-12}^{B} P_{-34}^{B}$ to measure the observable $W$. We illustrate these two measurement settings in Fig. \ref{0}(a) and Fig. \ref{0}(b), respectively.

For qubits, $P_-$ takes the particularly simple form $P_-=|\psi^-\rangle\langle\psi^-|$ with $|\psi^-\rangle=(|01\rangle-|10\rangle)/\sqrt{2}$, and it has already been measured in several experiments using two copies of $\varrho$. For example, there are at least two methods to measure $P_-$ in photonic system. The first method to project two photons onto the singlet state is to use a Hong-Ou-Mandel interferometer \cite{Ou}, which has been used in \cite{Huang}. The second method, employed in \cite{mintert}, is to distinguish the Bell states with a controlled-NOT gate, which can transform the Bell states to separable states.

Each of the two approaches proposed above to obtain the value of $\mathrm{Tr}(W\varrho^{\otimes 4})$,  namely by the quantum circuit and the direct measurement, has its own advantages. The quantum circuit requires only one local measurement setting, but it needs the help of four auxiliary qubits and four controlled-$U$ gates. The direct measurement of $W$ does not require any auxiliary qubit, however, it needs two measurement settings. Compared with the quantum circuit, the direct measurement of $W$ might be a little easier to realize in experiments, since the measurement of $P_-$ for qubits has been realized in experiments.

\section{Discussion and Conclusion}

In this work, we present a single observable $W$ to witness the quantum discord of an unknown state provided there are four copies of the state. The expectation value of our observable provides a necessary condition for nonzero quantum discord which turns out to be also sufficient in $2\times N$ systems. Also a nontrivial lower bound, which is illustrated in the example of DQC1, is placed by the discord witness. Moreover, a quantum circuit is designed to determine the expectation value of the discord witness $W$ using four simultaneous local qubit-measurements. Interestingly, four copies of states have also been turned out to be sufficient for entanglement detection of two-qubit states \cite{augusiak}.

\section*{ACKNOWLEDGMENTS}
This work is supported by National Research Foundation and Ministry of Education, Singapore (Grant No. WBS: R-710-000-008-271) and NNSF of China (Grant No. 11075227).

\section*{APPENDIX}
We will present the calculation details to support our results and statements in the main text.

\textbf{Complete proof of the Theorem.} Suppose that $\{A_i\}_{i=0}^{d^2_{A}-1}$ and $\{B_j\}_{j=0}^{d^2_{B}-1}$ are complete sets of LOOs for subsystem $A$ and $B$ \cite{LOO}, respectively, which satisfy $\mathrm{Tr}(A_i A_j)=\mathrm{Tr}(B_i B_j)=\delta_{ij}$ with $A_0=\idol^A/\sqrt{d_A}$ and $B_0=\idol^B/\sqrt{d_B}$.
As an example, one can choose $A_0=\idol^A/\sqrt{d_A}$, $B_0=\idol^B/\sqrt{d_B}$, $\{A_i\}_{i=1}^{d^2_{A}-1}$ and $\{B_j\}_{j=1}^{d^2_{B}-1}$ as the normalized traceless Hermitian generators of SU($d_A$) $\{\lambda_i\}_{i=1}^{d^2_{A}-1}$ and SU($d_B$) $\{\tilde{\lambda}_j\}_{j=1}^{d^2_{B}-1}$, respectively, i.e., $\tr\lambda_i=\tr\tilde{\lambda}_j=0$ and $\tr(\lambda_i\lambda_j)=\tr(\tilde{\lambda}_i\tilde{\lambda}_j)=\delta_{ij}$. Therefore any bipartite density matrix has expansion
\begin{widetext}
\begin{eqnarray}\label{a1}
\varrho=\frac{\idol^A}{d_A}\otimes\frac{\idol^B}{d_B}+\sum_{i=1}^{d_A^2-1}x_i\lambda_i\otimes\frac{\idol^B}{\sqrt{d_B}}+\sum_{j=1}^{d_B^2-1}y_j\frac{\idol^A}{\sqrt{d_A}}\otimes\tilde{\lambda}_j+\sum_{i=1}^{d_A^2-1}\sum_{j=1}^{d_B^2-1}t_{ij}\lambda_i\otimes\tilde{\lambda}_j,
\end{eqnarray}
\end{widetext}
where the coefficients $x_i$, $y_j$, $t_{ij}$ forms the column vectors $\mathbf{x}$, $\mathbf{y}$ and the $(d_A^2-1)\times(d_B^2-1)$ dimensional matrix $T$, respectively.

Denote by $R$ a $(d_A^2-1)\times(d_B^2)$ dimensional matrix whose matrix elements are $r_{ij}=\langle A_i\otimes B_j\rangle_{\varrho}$ with $i=1,\cdots,d_A^2-1$ and $j=0,\cdots,d_B^2-1$. For the state $\varrho$ given in Eq. (\ref{a1}), we have
\begin{equation}\label{}
    R=\left(\begin{array}{cc}
    \mathbf{x}_{(d_A^2-1)\times 1}| &  T_{(d_A^2-1)\times(d_B^2-1)}
   \end{array}\right).
\end{equation}
In general we have
\begin{eqnarray}\label{arho}
    \varrho&=&\sum_{i=0}^{d_A^2-1}\sum_{j=0}^{d_B^2-1}r_{ij}A_i\otimes B_j\nonumber\\
    &=&\sum_{i=1}^{d_A^2-1}\sum_{j=0}^{d_B^2-1}r_{ij}A_i\otimes B_j+\frac{\idol^A}{d_A}\otimes\varrho_B,
\end{eqnarray}
where $r_{0j}:=\langle A_0\otimes B_j\rangle_{\varrho}$ and we have used
\begin{eqnarray}\label{}
    &&\sum_{j=0}^{d_B^2-1}r_{0j}A_0\otimes B_j\nonumber\\
    &=&\sum_{j=0}^{d_B^2-1}r_{0j}\frac{\idol^A}{\sqrt{d_A}}\otimes B_j\nonumber\\
    &=&\sum_{j=0}^{d_B^2-1}\frac{1}{\sqrt{d_A}}\tr(\varrho_BB_j)\frac{\idol^A}{\sqrt{d_A}}\otimes B_j\nonumber\\
    &=&\frac{\idol^A}{d_A}\otimes\varrho_B,
\end{eqnarray}
since $r_{0j}=\langle \idol^A/\sqrt{d_A}\otimes B_j\rangle_{\varrho}=\tr(\varrho_BB_j)/\sqrt{d_A}$.
A straightforward calculation yields
\begin{eqnarray}\label{}
    &&\mathrm{Tr}(RR^T)\nonumber\\
    &=&\sum_{i=1}^{d_A^2-1}\sum_{j=0}^{d_B^2-1}r_{ij}r_{ij}\nonumber\\
    &=&\sum_{i=1}^{d_A^2-1}\sum_{j=0}^{d_B^2-1}\langle A_i^1\otimes B_j^1\rangle_{\varrho}\langle A_i^2\otimes B_j^2\rangle_{\varrho}\nonumber\\
    &=&\sum_{i=1}^{d_A^2-1}\sum_{j=0}^{d_B^2-1}\mathrm{Tr}\left( (A_i^1\otimes B_j^1\otimes A_i^2\otimes B_j^2) (\varrho^{\otimes 2})\right)\nonumber\\
    &=&\mathrm{Tr}(\bar{V}_{12}^A V_{12}^B \varrho^{\otimes 2}),
\end{eqnarray}
where $A_i^m$ and $B_j^n$ operate on the $m$th subsystem $A$ and the $n$th subsystem $B$, respectively. We have used the swap operator
\begin{equation}\label{av12B}
    V_{12}^B=\sum_{j=0}^{d_B^2-1}B_j^1\otimes B_j^2,
\end{equation}
which operates on the first and second copies of subsystem $B$, and defined
\begin{equation}\label{}
    \bar{V}_{12}^{A}:=\sum_{i=1}^{d_A^2-1}A_i^1\otimes A_i^2=V_{12}^{A}-\frac{\idol^{A}}{\sqrt{d_A}}\otimes\frac{\idol^{A}}{\sqrt{d_A}},
\end{equation}
operating on the first and second copies of subsystem $A$. Notice that the swap operator is defined as
\begin{equation}\label{}
    V:=\sum_{kl}|k\rangle\langle l|\otimes|l\rangle\langle k|.
\end{equation}
It is worth noticing that every Hermitian operator can be written in the operator basis, for example, the swap operator $V_{12}^{B}$ on the first and second copies of subsystem $B$
can be written as
\begin{equation}\label{adv}
    V_{12}^{B}=\sum_{i,j=0}^{d_B^2-1}c_{ij}B_i^1\otimes B_j^2,
\end{equation}
where
\begin{eqnarray}\label{acij}
c_{ij}&=&\tr(B_i^1\otimes B_j^2 V_{12}^{B})\nonumber\\
&=&\sum_{kl}\langle l|B_i^1|k\rangle\langle k|B_j^2|l\rangle\nonumber\\
&=&\tr(B_i^1 B_j^2)\nonumber\\
&=&\delta_{ij}.
\end{eqnarray}
Hence, one can directly obtain Eq. (\ref{av12B}) using Eqs. (\ref{adv}) and (\ref{acij}).
On the other hand, we have
\begin{eqnarray}\label{}
    &&\mathrm{Tr}[(RR^T)^2]\nonumber\\
    &=&\sum_{i,k=1}^{d_A^2-1}\sum_{j,l=0}^{d_B^2-1}\langle A_i^1\otimes B_j^1\rangle_{\varrho}\langle A_k^2\otimes B_j^2\rangle_{\varrho}\langle A_k^3\otimes B_l^3\rangle_{\varrho}\langle A_i^4\otimes B_l^4\rangle_{\varrho}\nonumber\\
    &=&\mathrm{Tr}(\bar{V}_{14}^{A}\bar{V}_{23}^{A}V_{12}^{B}V_{34}^{B}\varrho^{\otimes 4}).
\end{eqnarray}
Therefore,
\begin{eqnarray}\label{a9}
\mathrm{Tr}[(RR^T)^2]-[\mathrm{Tr}(RR^T)]^2=\mathrm{Tr}(W'\varrho^{\otimes4})=\mathrm{Tr}(W\varrho^{\otimes4}),
\end{eqnarray}
where we have defined
\begin{equation}\label{}
    W':=\bar{V}_{14}^{A}\bar{V}_{23}^{A}V_{12}^{B}V_{34}^{B}-\bar{V}_{12}^{A}\bar{V}_{34}^{A}V_{12}^{B}V_{34}^{B},
\end{equation}
and the second equality in Eq. (\ref{a9}) holds since
\begin{eqnarray}
&&\tr(V_{14}^AV_{12}^BV_{34}^B\varrho^{\otimes4})=\tr(V_{23}^AV_{12}^BV_{34}^B\varrho^{\otimes4}),\\
&&\tr(V_{12}^AV_{12}^BV_{34}^B\varrho^{\otimes4})=\tr(V_{34}^AV_{12}^BV_{34}^B\varrho^{\otimes4}),
\end{eqnarray}
which comes from the invariance of $\varrho^{\otimes4}$ under arbitrary permutation of different copies.

(i). If $\mathrm{Tr}(W\varrho^{\otimes4})=0$ then we have $\mathrm{Tr}[(RR^T)^2]=[\mathrm{Tr}(RR^T)]^2$, which means that the rank of the matrix $R$ is 0 or 1.

(ia) If $\rank(R)=0$, then $R=0$. From Eq. (\ref{arho}), we can obtain
\begin{equation}\label{}
    \varrho=\frac{\idol^A}{d_A}\otimes\varrho_B,
\end{equation}
which obviously has zero quantum discord $D_A(\varrho)=0$.

(ib) If $\rank(R)=1$, then there exist real numbers $\alpha_i$ and $\beta_j$ such that $r_{ij}=\alpha_i\beta_j$ with $i=1,\cdots,d_A^2-1$ and $j=0,\cdots,d_B^2-1$. It follows from Eq. (\ref{arho}) that
\begin{equation}\label{}
  \varrho=\sum_{i=1}^{d_A^2-1}\alpha_i A_i\otimes \sum_{j=0}^{d_B^2-1}\beta_j B_j+\frac{\idol^A}{d_A}\otimes\varrho_B,
\end{equation}
Using the eigenvectors $\{\Pi_k\}$ of $\sum_{i=1}^{d_A^2-1}\alpha_i A_i$ one can get $\sum_k(\Pi_k\otimes\idol)\varrho(\Pi_k\otimes\idol)=\varrho$, i.e., $D_A(\varrho)=0$.

Therefore, if $\mathrm{Tr}(W\varrho^{\otimes4})=0$ then we have $D_A(\varrho)=0$.

(ii). Conversely if $d_A=2$ and $D_A(\varrho)=0$ then there exists an orthonormal basis $\{|\psi_1\rangle,|\psi_2\rangle\}$ of subsystem $A$ and two density matrices $\varrho_k^B$ of subsystem $B$ such that
 \begin{equation}\label{a0}
   \varrho=p|\psi_1\rangle\langle\psi_1|\otimes\varrho_1^B+(1-p)|\psi_2\rangle\langle\psi_2|\otimes\varrho_2^B,
 \end{equation}
with $0\leq p\leq1$. Using Eq. (\ref{a0}), it is straightforward to check that $\mathrm{Tr}(W\varrho^{\otimes4})=0$. Therefore, for a $2\times N$ state $\varrho$, the discord $D_A(\varrho)=0$ if and only if $\mathrm{Tr}(W\varrho^{\otimes4})=0$.

Moreover, if we perform a von Neumann measurement $\{\Pi_k\}_{k=1}^{d_A}$ on the subsystem $A$
then we will obtain the outcome $k$ with probability $p_k=\mathrm{Tr}(\Pi_k\otimes\idol\varrho)$ leaving the subsystem $B$ in the state $\varrho_{B|k}=\varrho_{B|k}'/p_k$ where $\varrho_{B|k}'=\mathrm{Tr}_A(\Pi_k\otimes\idol\varrho)$. The conditional entropy becomes
\begin{eqnarray}
&&\sum_{k=1}^{d_A} p_k S(\varrho_{B|k})\nonumber\\
&=&\sum_{k=1}^{d_A} p_k S(\varrho_{B|k}'/p_k)\nonumber\\
&=&-\sum_{k=1}^{d_A} \tr[\varrho_{B|k}^\prime\log_2 (\varrho_{B|k}^\prime/p_k)]\nonumber\\
&=&\sum_{k=1}^{d_A} p_k\log_2 p_k-\sum_{k=1}^{d_A} \tr(\varrho_{B|k}^\prime\log_2 \varrho_{B|k}^\prime)\nonumber\\
&=&\sum_{k=1}^{d_A} p_k\log_2 p_k+S\left(\textstyle\sum_{k=1}^{d_A} \Pi_k\otimes\varrho_{B|k}'\right)\nonumber\\
&\geq&-\log_2 d_A-\log_2\big\{\textstyle\sum_{k=1}^{d_A}\tr[(\varrho_{B|k}')^2]\big\},
\end{eqnarray}
where  we have used the inequalities $\sum_{k=1}^{d_A} p_k\log_2 p_k\geq-\log_2 d_A$ and $S(\varrho)\ge -\log_2 \tr\varrho^2$. By expanding $\Pi_k$ using a complete set of LOOs with coefficients $e_i^k=\tr(\Pi_k A_i)$, i.e., $\Pi_k=\sum_{i=0}^{d_A^2-1} e_i^k A_i$,
and denoting by $M$ a $(d_A^2-1)\times(d_A^2-1)$ matrix with elements
$M_{ij}=\mathrm{Tr}(A_i^1\otimes A_j^2 V_{12}^{B}\varrho^{\otimes2})$, we obtain
\begin{eqnarray}
&&\sum_{k=1}^{d_A}\tr[(\varrho_{B|k}')^2]\nonumber\\
&=&\sum_{k=1}^{d_A}\tr(V_{12}^B\varrho_{B|k}'^{\otimes2})\nonumber\\
&=&\sum_{k=1}^{d_A}\sum_{i,j=0}^{d_A^2-1}e_i^k e_j^k \mathrm{Tr}(A_i^1\otimes A_j^2 V_{12}^{B}\varrho^{\otimes2})\nonumber\\
&=&\sum_{k=1}^{d_A}\sum_{i,j=1}^{d_A^2-1}e_i^k e_j^k M_{ij}+\frac{\mathrm{Tr}\varrho_B^2}{d_A}\nonumber\\
&\leq&\sum_{k=1}^{d_A}\sum_{i=1}^{d_A^2-1}(e_i^k)^2\sqrt{\mathrm{Tr}(MM^T)}+\frac{\mathrm{Tr}\varrho_B^2}{d_A}\nonumber\\
&=&(d_A-1)\sqrt{\mathrm{Tr}(W\varrho^{\otimes4})+(\tr\tilde\varrho^2)^2}+\frac{\mathrm{Tr}\varrho_B^2}{d_A}\nonumber\\
&:=&Q(\varrho),\label{a15}
\end{eqnarray}
where we have used the Cauchy-Schwarz inequality, $\sum_{i=1}^{d_A^2-1}(e_i^k)^2=(d_A-1)/d_A$, and $\sum_{k=1}^{d_A}e_i^k=0$ for $i=1,\cdots,d_A^2-1$. Therefore,
\begin{eqnarray}
&&D_A(\varrho)\nonumber\\
&=&S(\varrho_A)-S(\varrho)+\min_{\{E_k\}}\textstyle\sum_k p_k S(\varrho_{B|k})\nonumber\\
&\geq&S(\varrho_A)-S(\varrho)-\log_2 d_A-\log_2\big\{\textstyle\sum_{k=1}^{d_A}\tr[(\varrho_{B|k}')^2]\big\}\nonumber\\
&\geq&S(\varrho_A)-S(\varrho)-\log_2 d_A-\log_2[Q(\varrho)]\nonumber\\
&=&S(\varrho_A)-S(\varrho)-\log_2 [d_AQ(\varrho)],\label{alower}
\end{eqnarray}
which is the lower bound of the quantum discord Eq. (\ref{lower}) in the main text.

\textbf{Remark.}
Similarly to Eq. (\ref{alower}), one can also present a nontrivial lower bound for the geometric measure of quantum discord $D_A^{(2)}(\varrho)$ using the expectation value of $W$,
where $D_A^{(2)}(\varrho)$ is the geometric measure of quantum discord defined in Ref. \cite{condition1}.
Actually, Ref. \cite{luo2} has shown that
\begin{equation}\label{}
    D_A^{(2)}(\varrho)=\min_{\{\Pi_k\}}\left\{\textstyle\tr\varrho^2-\sum_{k}\tr(\varrho_{B|k}')^2\right\},
\end{equation}
Using Eq. (\ref{a15}), one can obtain
\begin{eqnarray}
 D_A^{(2)}(\varrho)&=&\min_{\{\Pi_k\}}\left\{\textstyle\tr\varrho^2-\sum_{k}\tr(\varrho_{B|k}')^2\right\}\nonumber\\
 &=&\tr\varrho^2-\max_{\{\Pi_k\}}\left\{\textstyle\sum_{k}\tr(\varrho_{B|k}')^2\right\}\nonumber\\
 &\geq&\tr\varrho^2-Q(\varrho),
\end{eqnarray}
which is Eq. (\ref{lower2}) in the main text.

\textbf{DQC1 Model.} The output state of DQC1 model can be written as
\begin{eqnarray}\label{afinal}
    \varrho_{out}&=&\frac{1}{2^{n+1}}
\left(\begin{array}{cc}
 \idol_n & \alpha U_{n}^{\dag} \\
 \alpha U_{n} &  \idol_n
\end{array}\right)\nonumber\\
    &=&\frac{1}{2^{n+1}}(\idol_1\otimes\idol_n+\alpha|1\rangle\langle0|\otimes U_n+\alpha|0\rangle\langle1|\otimes U_n^{\dag})\nonumber\\
    &=&\frac{1}{2^{n+1}}\sum_{i,j=0}^1\alpha^{(i+j)\mathrm{mod}2}|i\rangle\langle j|\otimes U_n^i {U_n^{\dag}}^{j}.
\end{eqnarray}
Notice that for DQC1 model $d_A=2$, thus
\begin{equation}\label{}
   W=U_1+U_4-U_2-U_3.
\end{equation}
Based on Eq. (\ref{afinal}), one can directly calculate as follows,
\begin{eqnarray}
\tr(U_1\varrho_{out}^{\otimes4})&=&\frac{1}{2^{4n+4}}\big(2^{2n+2}+8\alpha^2|\tr U_n|^2\nonumber\\
&&+2\alpha^4|\tr U_n^2|^2+2^{2n+1}\alpha^4\big),\\
\tr(U_2\varrho_{out}^{\otimes4})&=&\frac{1}{2^{4n+4}}\big(2^{2n+3}+8\alpha^2|\tr U_n|^2\big),\\
\tr(U_3\varrho_{out}^{\otimes4})&=&\frac{1}{2^{4n+4}}\big(2^{2n+2}+2^{2n+3}\alpha^2\nonumber\\
&&+2^{2n+2}\alpha^4\big),\\
\tr(U_4\varrho_{out}^{\otimes4})&=&\frac{1}{2^{4n+4}}\big(2^{2n+3}+2^{2n+3}\alpha^2\big).
\end{eqnarray}
Therefore,
\begin{eqnarray}\label{}
    \tr(W\varrho_{out}^{\otimes4})&=&\tr[(U_1+U_4-U_2-U_3)\varrho_{out}^{\otimes4})]\nonumber\\
    &=&\frac{\alpha^4}{2^{2n+3}}\bigg(\bigg|\frac{\mathrm{Tr}U_n^2}{2^n}\bigg|^2-1\bigg),
\end{eqnarray}
which is Eq. (\ref{wdqc1}) in the main text. According to our theorem, $D_A(\varrho_{out})=0$ if and only if $\mathrm{Tr}(W\varrho_{out}^{\otimes4})=0$, i.e., $\alpha=0$ or $U_n^2=e^{i\phi}\idol_n$ for some real $\phi$, which is equivalent to the result shown in Ref. \cite{condition1}. It has been shown that for almost every unitary matrix $U_n$ (random unitary) the discord in the output state (\ref{afinal}) is nonvanishing \cite{Datta}. However, in the nontrivial case $U_n^2=e^{i\phi}\idol_n$ the quantum discord $D_A(\varrho_{out})=0$, which provides a counter example for the nonvanishing quantum discord of every nontrivial unitary matrix $U_n$. Thus, the quantum discord might not be the reason for the quantum speedup in DQC1.

On the other hand, if $\mathrm{Tr}U_n=0$ and $\alpha=1$, then for the output state Eq. (\ref{afinal}) one can obtain that $\varrho_A=\idol_1/2$, $\varrho_B=\idol_n/2^n$, and the eigenvalues of $\varrho_{out}$ are $\{0,\cdots,0,1/2^n,\cdots,1/2^n\}$ in which there are $2^n$ copies of $0$ and $2^n$ copies of $1/2^n$. Therefore,
\begin{eqnarray}
&&S(\varrho_A)=1,\\
&&S(\varrho_{out})=n,\\
&&\tr\varrho_B^2=\frac{1}{2^n},\\
&&\tr\tilde{\varrho}_{out}^2=\frac{1}{2^{n+1}}.
\end{eqnarray}
Using Eq. (\ref{alower}) we have a lower bound for the quantum discord in DQC1
\begin{equation}\label{}
    D_A(\varrho_{out})\geq 1- \log_2\bigg(1+\sqrt{2^{2n+2}\mathrm{Tr}(W\varrho_{out}^{\otimes4})+1}\bigg),
\end{equation}
which is Eq. (\ref{dqc1lower}) in the main text. It is obvious that the negative expectation value $\mathrm{Tr}(W\varrho_{out}^{\otimes4})$ provides a nontrivial lower bound for $D_A(\varrho_{out})$.

\textbf{Measuring the discord witness $W$.}
In order to measure the expectation value of the observable $W$ we have designed a quantum circuit as shown in Fig. \ref{1} in the main text. We have introduced four auxiliary qubits, which are all initially prepared in the state $|+\rangle=(|0\rangle+|1\rangle)/\sqrt{2}$. After the controlled-$U_i$ ($i=1,\cdots,4$) operations with each qubit as source, we preform $\sigma_x$-measurements on all four auxiliary qubits so that four expectation values are obtained simultaneously. As a result, the total state including the four auxiliary qubits is
\begin{eqnarray}
\varrho_{total}&=&\frac{1}{16}\sum_{i,i',\cdots,l'}\Big(|i\rangle\langle i'|\otimes|j\rangle\langle j'|\otimes|k\rangle\langle k'|\otimes|l\rangle\langle l'|\nonumber\\
&&\otimes U_4^l U_3^k U_2^j U_1^i \varrho_{out}^{\otimes4}{U_1^{\dag}}^{i'}{U_2^{\dag}}^{j'}{U_3^{\dag}}^{k'}{U_4^{\dag}}^{l'}\Big).
\end{eqnarray}
Thus, we have
\begin{eqnarray}
\langle \sigma_x^1\rangle&=&\frac{1}{2}[\mathrm{Tr}(\varrho^{\otimes4}U_1^{\dag})+\mathrm{Tr}(U_1\varrho^{\otimes4})]\nonumber\\
&=&\mathrm{Tr}(U_1\varrho^{\otimes4}),\\
\langle \sigma_x^2\rangle&=&\frac{1}{2}[\mathrm{Tr}(\varrho^{\otimes4}U_2^{\dag})+\mathrm{Tr}(U_2\varrho^{\otimes4})]\nonumber\\
&=&\mathrm{Tr}(U_2\varrho^{\otimes4}),\\
\langle \sigma_x^3\rangle&=&\frac14[\mathrm{Tr}(\varrho^{\otimes4}U_3^{\dag})+\mathrm{Tr}(U_3\varrho^{\otimes4})+2\langle \sigma_x^1\rangle]\nonumber\\
&=&\frac12[\mathrm{Tr}(U_3\varrho^{\otimes4})+\langle \sigma_x^1\rangle],\\
\langle \sigma_x^4\rangle&=&\frac14[\mathrm{Tr}(\varrho^{\otimes4}U_4^{\dag})+\mathrm{Tr}(U_4\varrho^{\otimes4})+2\langle \sigma_x^2\rangle]\nonumber\\
&=&\frac12[\mathrm{Tr}(U_4\varrho^{\otimes4})+\langle \sigma_x^2\rangle],
\end{eqnarray}
from which one can easily calculate the value of $\mathrm{Tr}(W\varrho^{\otimes 4})$ according to
\begin{equation}\label{asigma}
    \mathrm{Tr}(W\varrho^{\otimes 4})=2\Big(\langle \sigma_x^1\rangle-\langle \sigma_x^3\rangle\Big)-\frac{4}{d_A}\Big(\langle \sigma_x^2\rangle-\langle \sigma_x^4\rangle\Big),
\end{equation}
which is Eq.(\ref{sigma}) in the main text. In order to obtain Eq. (\ref{P_W}) in the main text, one can use the definition of $W$ Eqs. (\ref{U}) and (\ref{W}) in the main text and the equation $V=\idol-2P_-$.

\end{document}